# Magnetoresistance of individual ferromagnetic GaAs/(Ga,Mn)As core-shell nanowires

*Christian H. Butschkow,*[†] *Elisabeth Reiger,*[†] *Stefan Geißler,*[†] *Andreas Rudolph,*[†] *Marcello Soda,*[†] *Dieter Schuh,*[†] *Georg Woltersdorf,*[†] *Werner Wegscheider,*[†,¶] *and Dieter Weiss*[†]

[†] Institute for Experimental and Applied Physics, University of Regensburg, D-93040 Regensburg, Germany

[¶] ETH Zürich, 8093 Zürich, Switzerland

E-mail: christian.butschkow@physik.uni-regensburg.de

We investigate, angle dependent, the magnetoresistance (MR) of individual self-assembled ferromagnetic GaAs/(Ga,Mn)As core-shell nanowires at cryogenic temperatures. The shape of the MR-traces and the observed strong anisotropies in transport can be ascribed to the interplay of the negative magnetoresistance effect and a strong uniaxial anisotropy with the magnetic easy direction pointing along the wire axis. The magnetoresistance can be well described by a quantitative analysis based on the concept of the effective magnetic field, usually used to describe ferromagnetic resonance phenomena. The nanowires we investigate exhibit a uniaxial anisotropy which is approximately 5 times larger than the strain induced anisotropy observed in lithographically prepared (Ga,Mn)As stripes.

Keywords: (Ga,Mn)As, nanowire, ferromagnetism, core-shell, magnetotransport, NMR

Self-assembled ferromagnetic nanowires constitute a new and widely uncharted facet of semiconductor spintronics. The flexibility in the choice of material as well as the axial and radial degrees of freedom during self-assembly[1-3] make nanowires an interesting building block for nanoscale spintronic elements like one-dimensional spin valve transistors or ferromagnetic single electron transistors.[4,5] For the latter a ferromagnetic (non-magnetic) wire-segment needs to be incorporated into a non-magnetic (magnetic) wire segment. The best characterized and understood magnetic semiconductor is to date (Ga,Mn)As so that it is tempting to investigate the properties of this paradigmatic ferromagnetic semiconductor in a self-assembled nanowire geometry. In bulk (Ga,Mn)As the ferromagnetic interaction between the Mn ions with spin 5/2 is mediated by the mobile holes via the RKKY interaction.[6] (Ga,Mn)As needs to be grown at low temperatures to avoid segregation of Mn-clusters although growth of (Ga,Mn)As nanowires at elevated temperature has been reported.[7,8] To stay at low growth temperatures we resort to the core-shell approach, i.e. a GaAs core wire is assembled first and then, at lower growth temperature, a ferromagnetic (Ga,Mn)As shell is added. This constitutes an interesting topology, a non-magnetic GaAs core surrounded by a ferromagnetic cylinder.

While there already exist a number of publications on the synthesis of different dilute magnetic II-VI, III-V and IV[9-12] semiconductor nanowires, the properties of MgO/$Fe_3O_4$,[13] GaAs/MnAs[14,15] and GaAs/(Ga,Mn)As[16] self-assembled core-shell wires or magnetotransport properties of GaN:Mn[10], ZnO:Co[17] or MnSi[18] nanowires, a detailed magnetotransport study on the topologically different ferromagnetic core-shell GaAs/(Ga,Mn)As nanowires is still lacking. The geometry of the wire gives rise to a strong uniaxial magnetic anisotropy with anisotropy constants strongly exceeding the one of corresponding planar wires of the same material. Since the magnetic anisotropies in (Ga,Mn)As layers are essentially determined by the built-in strain,[19] we speculate that the strong uniaxial anisotropy is due to the peculiar strain pattern and strain relaxation at the cylindrical GaAs/(Ga,Mn)As heterointerface.

Because the growth of the GaAs/(Ga,Mn)As core-shell nanowires and some of their (ensemble) properties were described previously[16] we summarize only some basic features here. The GaAs core nanowires were grown at a substrate temperature of 533° C via the vapor-liquid-solid mechanism by molecular beam epitaxy using gold as catalyst. The diameters of the resulting nanowires range between 40 nm and 80 nm, with lengths between 500 nm and 4.5 µm. All nanowires grew perpendicularly on the (111)B substrate wafer. The (Ga,Mn)As shell was grown at a reduced temperature of 205° C. The amount of deposited (Ga,Mn)As corresponds to a 200 nm thick layer, containing approximately 5% Mn. If (Ga,Mn)As was homogeneously distributed over the whole NW surface area, this would result in an average shell thickness of ~20 nm. Shadowing effects by the neighboring NWs lead to a thinning of the shell towards the substrate, visible in figure 1a) and often to a radially asymmetric shell growth, as observed in the cross-section displayed in figure 1b). In bright field TEM images of a nanowire cross section (c.f. figure 1b)) the GaAs core can be clearly distinguished from the shell. The contrast stems from the Mn interstitials in (Ga,Mn)As,[20] indicating that no pronounced Mn back-diffusion into the core takes place. This is also supported by energy dispersive X-ray spectroscopy (EDX) measurements (not shown). Therefore we do not expect a distinct parallel conduction in core and shell, which might influence the magnetotransport measurements.

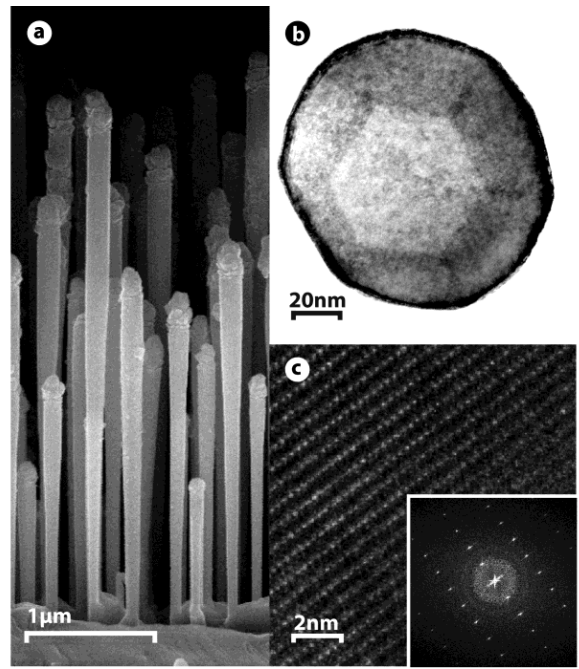

**Figure 1.a)** Side-view scanning electron micrograph (SEM) of an as-grown nanowire sample. The (Ga,Mn)As shell gets thinner towards the substrate leading to a tapered shape of the nanowires. b) Bright-field transmission electron micrograph (TEM) of a core-shell nanowire cross section. The hexagonal core of GaAs with $\{11\bar{2}0\}$ oriented side facets is clearly distinguishable from the (Ga,Mn)As shell. c) Side-view high-resolution TEM micrograph of the middle part of a core nanowire and its diffractogram (inset). The crystal structure is pure wurtzite.

The absence of crystal defects in the side-view of the shell indicates the successful epitaxial growth of (Ga,Mn)As on the core nanowires. The crystal structure depicted in the HRTEM side-view micrograph is wurtzite as can be recognized from the diffractogram of figure 1c). Indeed, in the central section of the nanowires (figure 1c)) which we probe in transport experiments, the crystal structure is pure wurtzite while almost all stacking faults occur at the top of the wires.

In a previous study (Ga,Mn)As core-shell NWs were characterized by SQUID (superconducting quantum interference device) magnetometry.[16] These experiments carried out on a 5 x 5 mm² piece of as-grown wafer with a large ensemble of nanowires revealed a ferromagnetic transition at 20 K and a uniaxial magnetic anisotropy with a magnetic easy axis pointing along the nanowire axis.

Here we probe electric and magnetic properties of individual nanowires employing magnetotransport. To isolate single wires the as-grown nanowires on the wafer were first sonicated into propanol. By transferring a droplet of the solution onto a Si/SiO2 substrate and by evaporating the propanol on a hot plate single wires get homogeneously distributed over the surface. Up to six contacts (cf. fig 2b), each with a width of 300 nm and a spacing of 400 nm were defined using electron beam lithography (EBL) and lift-off technique. After removing the native oxide by an HCl-dip followed by argon sputter cleaning inside the evaporation chamber, 15 nm Ti as an adhesion layer and 200nm Au were deposited and form ohmic contacts to the wire.

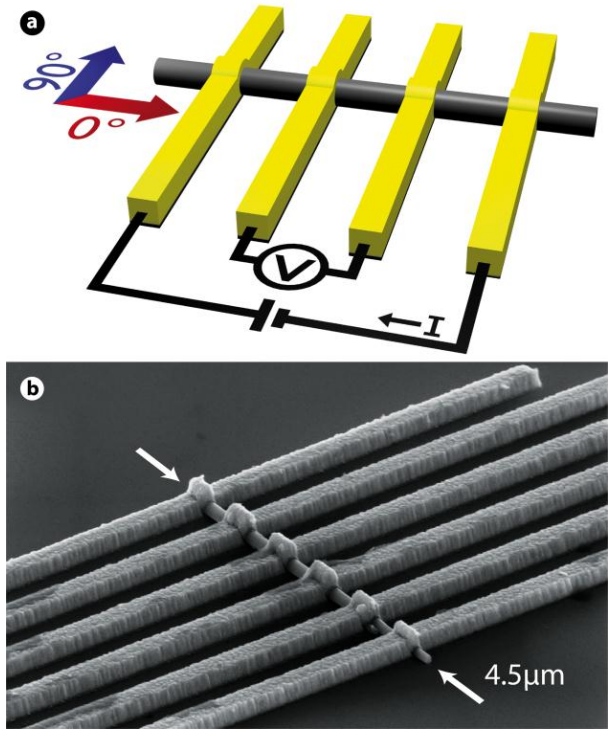

**Figure 2. a)** Schematic of the sample design and the experimental setup. The red and blue arrows span the sample plane. The directions 0°, pointing along the NW axis, and 90° oriented perpendicular to the NW axis are used in the text to give the direction of an external applied magnetic field. b) Tilted scanning electron micrograph (SEM) of a contacted nanowire.

Unless stated otherwise the transport measurements were performed in a 4-terminal configuration (cf. fig 2a), and carried out in a cryostat equipped with a superconducting coil enabling measurements at temperatures down to 1.4 K and in magnetic fields up to 10 T. The sample was mounted on a rotatable sample holder, permitting to vary the angle between applied magnetic field and nanowire axis. The Curie temperature of a single wire was extracted from the resistance $R(T)$: Following Novák et al.,[21] the actual Curie-Temperature is given by the singularity in $dR/dT$, the first derivative of the resistance curve during cool down. This way we determined the Curie-Temperature of five nanowires to be between 17 K and 19 K, in good agreement with the results of SQUID-measurements (The measurements of the Curie-Temperature via SQUID and magnetotransport are provided in the supporting information).

In total we investigated seven core-shell nanowires, which differ slightly in diameter and length but all show similar results. Therefore we focus in the following on the measurements on two nanowires. Typical magnetoresistance traces of one of these nanowires for the in-plane magnetic field $\vec{H}$ aligned in different directions are shown in figure 3. For high fields the negative magnetoresistance (NMR) dominates, but is by an order of magnitude larger than in planar (Ga,Mn)As films or wires.[22] The NMR is ascribed to electron-spin scattering. In particular, magnetic excitations like spin fluctuations or spin waves can be suppressed by an applied magnetic field which reduces electron scattering and thus the resistance.[23] Alternative explanations of the NMR effect in (Ga,Mn)As are based on weak localization in bulk material.[19]

For a magnetic field applied parallel to the wire axis (0°, red curve in figure 3) and sweeping the field up, $R$ displays a large resistance jump once the magnetization has changed polarity. In contrast, sweeps with a perpendicular magnetic field orientation (90°, blue curve in figure 3) exhibit a continuous change of the low field resistance. This is consistent with previous studies on nanowire ensembles, indicating a uniaxial magnetic anisotropy with the easy axis pointing along the nanowire axis (0°). The resistance jumps, can then be ascribed to an abrupt magnetization reversal due to thermally activated domain wall nucleation and propagation, characteristic for magnetization reversal along a magnetic easy axis. The continuous resistance change for the perpendicular configuration reflects a coherent rotation of the magnetization, characteristic for a magnetic hard axis. The grey curves in figure 3 illustrate the crossover from a parallel (0°) to a perpendicular (90°) field configuration. The magnetization reversal is then a superposition of coherent rotation and domain wall propagation.

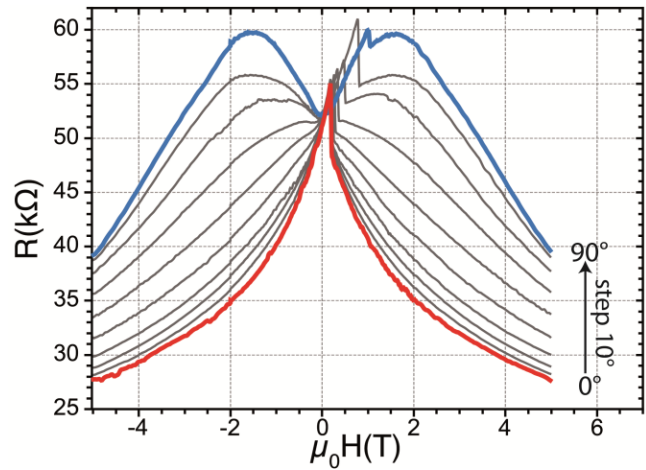

**Figure 3.** Magnetoresistance curves at T=4 K, measured in 2-terminal configuration, taken for different in-plane magnetic field directions between 0° (red) and 90° (blue) in steps of 10°. The field was swept from negative to positive values of $H_0$ (up sweep). The grey traces illustrate the gradual transition from the 0° to the 90° magnetoresistance trace. The small peak around ~1 T in the topmost trace indicates that the field was not perfectly aligned along the 90° direction (accuracy of alignment: +-2°).

A magnification of the low field magnetoresistance for both, up and down sweep of another wire are shown in figure 4. Also this sample shows the pronounced jumps for H parallel to the wire axis and a positive magnetoresistance which turns to a negative one for larger H, applied perpendicular to the wire axis. The nearly linear slope of $R(H)$ in figure 4a suggests that the NMR exhibits an approximately linear dependence on the applied H in this configuration for magnetic fields below 1 T.

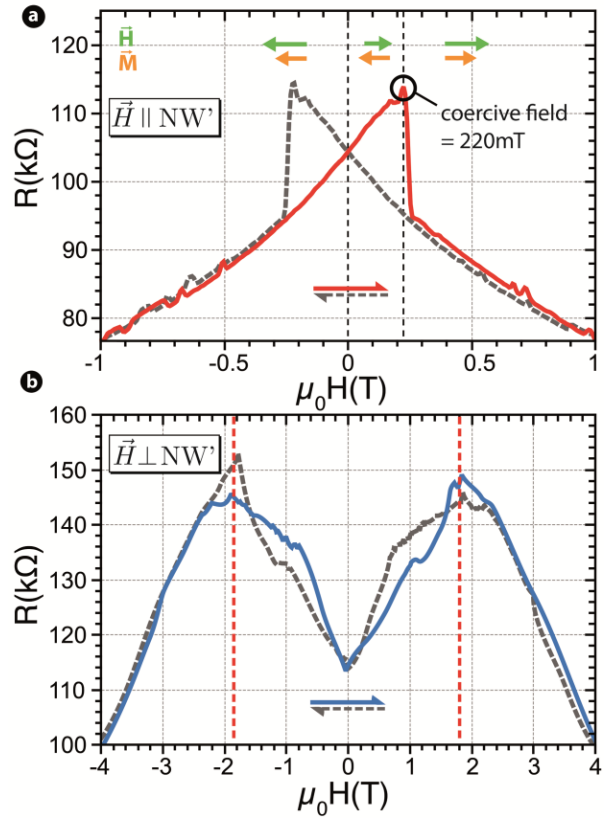

**Figure 4. a)** Magnetotransport data at T=4 K for $\vec{H}$ parallel to the nanowire axis (0°). The solid red line and the dashed grey line refer to opposite sweep directions. The colored arrows at the top represent qualitatively the direction and value of the external magnetic field $\vec{H}$ and the magnetization $\vec{M}$. **b)** Magnetic field sweep perpendicular to the nanowire axis (90°) at T=4 K. The two red dashed lines at the local maxima of the resistance curve mark the anisotropy field $H_a$. The resistance difference of ~10kΩ between sweeps in parallel and perpendicular configuration at zero magnetic field is ascribed to the splitting into several domains when the magnetic field is (at high B) perfectly aligned perpendicular to the easy axis and then decreased.

We now show that the dominating features observed in the magnetoresistance traces of figure 3 and 4 can be described by the dependence of the NMR on the effective magnetic field $\vec{H}_{eff}$. In general, the precession of the magnetization $\vec{M}$ can be described by the equation of motion $d\vec{M}/dt = -\gamma\left[\vec{M}\times\vec{H}_{eff}\right]$ with the gyromagnetic factor $\gamma$ and the effective field $\vec{H}_{eff}$. Precession takes place around the direction of the effective field $\vec{H}_{eff}$ which is parallel to the equilibrium position of the static magnetization.[24] It can be shown, that for small displacements of the magnetization, $\vec{H}_{eff}$ is defined by the Gauss curvature of the free energy with respect to the polar and azimuthal angle of the magnetization vector $\vec{M}$. In our case, only a uniaxial anisotropy term with anisotropy constant $K_U$ and the Zeeman energy term need to be considered for the free energy:

$$E(\theta_M, \varphi_M) = K_U \sin^2(\theta_M)\sin^2(\varphi_M) + \mu_0 M_S H \sin(\theta_M)\cos(\varphi_M - \varphi_H) \quad (1)$$

$\theta_M$ and $\varphi_M$ specify the azimuthal and polar angles of the magnetization with respect to the magnetic easy axis, while $\varphi_H$ is the direction of the external magnetic field and $M_S$ the saturation magnetization. When the static magnetization is oriented in the plane of the substrate ($\theta_M = 90°$) equation (1) leads to the expression for the effective field:

$$H_{eff} = \frac{1}{\mu_0 M_S \sin(\theta_M)}\sqrt{E_{\theta_M \theta_M} E_{\varphi_M \varphi_M} - E^2_{\theta_M \varphi_M}} \stackrel{\theta_M = 90°}{\approx} H_0 \cos(\varphi_M - \varphi_H) + \frac{2K_U}{\mu_0 M_S}\cos(2\varphi_M) \quad (2)$$

where $E_{\theta_M \theta_M}, E_{\varphi_M \varphi_M}$ and $E_{\theta_M \varphi_M}$ are the second derivatives of the free energy landscape.

The effective field consists of two terms, the first term is related to the applied external magnetic field and the second term to the anisotropy field $H_a = \frac{2K_U}{\mu_0 M_S}$.

For $\vec{H}$ applied parallel to the wire the anisotropy field points in the same direction as the applied field. The observed linear dependence of $R(H)$ in the parallel field configuration, displayed in Fig. 4a, suggests that the NMR changes linearly with the effective field. Therefore we assume in the following that the NMR changes, at least to first order, linearly with the effective field $H_{eff}$. In such a picture the measured resistance depends strongly on $H_a$ and is large for small effective field and small for large effective field.

Let us analyze the data of figure 4 within such a picture. The magnetoresistance measured along the nanowire axis (red graph fig. 4a) shows a nearly linear increase from -1 T to 0 T and continues to increase beyond $\mu_0 H = 0$ T, until a resistance jump occurs at the coercive field $H_C$. The values of $\mu_0 H_C$ obtained for different nanowires at 4 K range between 140 mT and 220 mT, in good agreement with SQUID data of nanowire ensembles. Prior to the magnetization reversal, the magnetization and the applied magnetic field, sketched in the inset of Fig. 4a, are pointing into opposite directions ($\varphi_M = 180°, \varphi_H = 0°$). According to equation (2) this leads for $H_0 = H_C$ to $H_{eff,1} = -H_C + \frac{2K_U}{\mu_0 M_S}$. After the jump, the magnetization aligns with the external field ($\varphi_M = 0°, \varphi_H = 0°$) and we obtain $H_{eff,2} = H_C + \frac{2K_U}{\mu_0 M_S}$. Thus, the resistance jump due to NMR reflects the absolute difference of the effective magnetic field values of $\Delta H_{eff} = 2H_C$. The corresponding backsweep (dashed gray curve) provides, due to ferromagnetic hysteresis, qualitatively the same measurement but mirrored at $\mu_0 H_0 = 0$ T.

In contrast, the magnetoresistance for $\vec{H}$ applied perpendicular to the wire axis (cf. figure 4b) changes almost continuously and exhibits only little hysteresis. Assuming a continuous rotation of the magnetization, the situation is also well described by equation (2). The two local maxima at ±1.8 T in figure 4b) mark approximately the anisotropy field $H_a$ where the magnetization saturates ($\varphi_M = 90°, \varphi_H = 90°$) and the effective field vanishes. I.e. the externally applied field equals the anisotropy field, and the effective magnetic field is zero. We observed values for $\mu_0 H_a$ in the range of 1.3 T to 2.2 T. From SQUID measurements we obtain $M_S \approx 9400 \frac{A}{M} \pm 2000 \frac{A}{M}$. Using $H_a = \frac{2K_U}{\mu_0 M_S}$ leads to $K_U \approx 8500 \frac{J}{m^3} \pm 1800 \frac{J}{m^3}$ for the magnetic anisotropy constant. (Footnote: The large error is a consequence of the uncertainty of the magnetically active sample volume for the nanowire array ensemble. The volume was estimated using the nanowire density determined by evaluating SEM images.)

The angular dependence of the magnetoresistance can be best visualized by monitoring the resistance during rotation of the sample in a fixed magnetic field. In figure 5 (purple curve) the resistance as a function of $\varphi_H$ for $\mu_0 H_a = 3\,\text{T}$ is shown. Due to the strong anisotropy the direction of the applied field $\varphi_H$ and the direction of the magnetization $\varphi_M$ differ quite significantly. Only when the applied field points along the magnetic easy or hard axis, the two directions coincide. Otherwise one observes the typical dragging of the magnetization behind the applied field. The direction of the magnetization can be calculated for each field angle by minimizing the free energy density (equation 2) ($\frac{\partial E}{\partial \varphi_M} = 0$ and $\frac{\partial^2 E}{\partial \varphi_M^2} > 0$). This procedure allows a plotting of the resistance values as a function of $\varphi_M$ (figure 5 green dots). For high magnetic fields $H_0 > H_a$ one finds $\cos(\varphi_M - \varphi_H) \approx 1$ and the angular dependence of the effective magnetic field becomes approximately $H_{\text{eff}} \approx H_0 + \frac{2K_U}{\mu_0 M_s}\cos(2\varphi_M)$. In figure 5 $R(\varphi_M)$ can be well fitted by $\cos(2\varphi_M)$, showing that the effective magnetic field dominates the magnetotransport behavior.

The small differences between data and the $\cos(2\varphi_M)$ fit are expected to reflect deviations from the supposed linear relationship between the measured resistance and $H_{\text{eff}}$, valid only to first order.

We note that the anisotropic magnetoresistance effect (AMR), which is also observed in (Ga,Mn)As nano-structures, shows the same angular dependence.[25] The AMR effect originates from spin-orbit coupling and depends mainly on the angle between current and magnetization ($\text{AMR} = \frac{R(\varphi_M) - R(0)}{R(0)} \propto \cos(2\varphi_M)$). In figure 5 we can thus not distinguish between NMR and AMR effect. However, the typical amplitude of the AMR effect is below 10 %,[25] which would constitute only a minor contribution to the observed MR effect of 74 %.

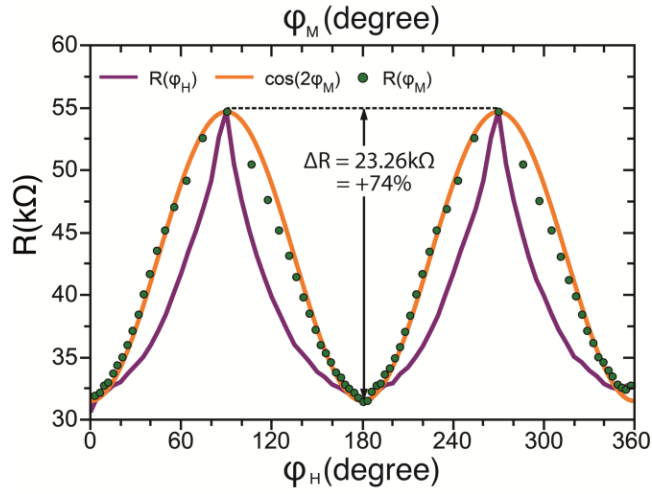

**Figure 5.** Angle dependence of the magnetoresistance at $\mu_0 H_0 = 3\text{T}$: The purple graph shows $R(\varphi_H)$ at $T = 4\text{K}$ while the green dots show $R(\varphi_M)$. $\varphi_M$ is obtained by minimizing the free energy equation for each data point. $R(\varphi_M)$ can be well fitted by $\cos(2\varphi_M)$ (orange curve) which describes approximately the angular dependence of the effective magnetic field.

For direct comparison of the magnetoresistance for parallel and perpendicular field configuration, the curves have to be displayed as a function of the effective magnetic field. For $\varphi_H = 0°$ and $H_0 > H_C$, $H_{\text{eff}} = H_0 + H_a$ holds while for $\varphi_H = 90°$ and $H_0 > H_a$ the effective magnetic field is $H_{\text{eff}} = H_0 - H_a$. By shifting both curves by $\mu_0 H_a$ along the x-axis to the right and to the left, respectively (figure 6 dashed lines), the two data sets can be compared directly on the effective magnetic field scale for $H_{\text{eff}} > H_a + H_C$ (figure 6 inset). While on the $\mu_0 H_0$ scale the magnetic field dependence of the resistance for parallel and perpendicularly applied appear to be rather different, they are nearly the same on the $\mu_0 H_{\text{eff}}$ scale, thus emphasizing the dominating role of the effective magnetic field for magnetotransport. The remaining difference is of the order of 3 % to 10 % and stems most likely from the AMR effect. The AMR effect usually does not show any field dependence. However, the resistance difference $R(90°) - R(0°)$ in the inset of figure 6 is not constant, suggesting that additional effects influence the magnetoresistance. For example weak localization[19] or, due to the inhomogeneity of the nanowire, spatially altering magnetic properties could contribute to the measurements.

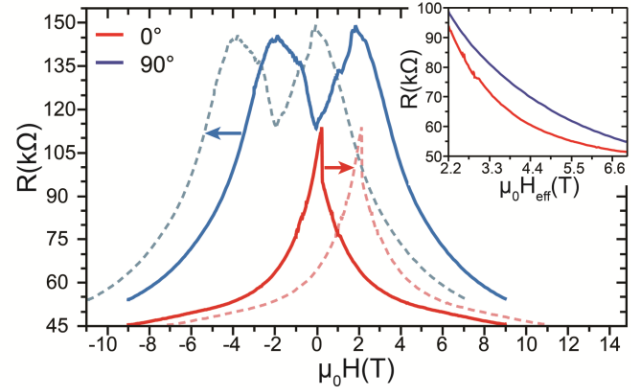

**Figure 6:** Illustration of the transformation from the $\mu_0 H_0$ scale to the $\mu_0 H_{eff}$ by shifting $R(90°)$ and $R(0°)$ by $\pm \mu_0 H_a$. The resulting high field magnetoresistance at 4 K as a function of the effective magnetic field is shown in the inset. This makes the two curves directly comparable for values of the effective field of $\mu_0 H_{eff} > 2.2$ T (inset).

The anisotropy field of typically ~2 T measured in our core-shell nanowires is nearly an order of magnitude larger than the one observed in wires made lithographically from (Ga,Mn)As planar films which range from $\mu_0 H_{eff}$ ~180 mT to 300 mT.[26,27] These lithographically fabricated nanowires show just as the core-shell nanowires a single magnetic easy axis coinciding with the wire axis. This is ascribed to a symmetry breaking, caused by the relaxation of the crystal lattice perpendicular to the stripe direction, which leads to a uniaxially strained crystal lattice.[28] We determined a rather low saturation magnetization of $\mu_0 M_S \approx 11.8$ mT for our samples. Typically high quality (Ga,Mn)As has an about three times larger saturation magnetization of ~37 mT.[29] We note, however, that also the uniaxial anisotropy constant $K_U \approx 8500$ J/m$^3$ is very high, when compared to etched nanowires, for which $K_U$ usually does not exceed 1500 J/m$^3$.[26] This might be a consequence of the peculiar strain and strain-relaxation pattern associated with the core-shell geometry. It is well known that strain crucially influences magnetic anisotropies.[19] In our sample geometry a sheet of (Ga,Mn)As is "wrapped" around a GaAs core with smaller lattice constant. The associated strain might be very effectively relaxed with increasing radius of the shell giving rise to the observed easy direction along the wire axis. Since strain effects occur primarily at the (Ga,Mn)As / GaAs interface with an homogeneously thick (Ga,Mn)As core we expect that the role of the thickness variations of the (Ga,Mn)As shell are of minor importance for $K_U$. Note that shape anisotropy only plays a minor role in (Ga,Mn)As due to the low saturation magnetization.[30] The fact that the central part of the nanowires has wurtzite crystal structure may also be responsible for the observed behavior. To analyze this, a more detailed knowledge of the (Ga,Mn)As/GaAs(wurtzite) interface and the connection between strain and magnetic anisotropy at this interface is required.

In conclusion we have performed magnetotransport measurements on individually contacted GaAs/(Ga,Mn)As core-shell nanowires which show a very pronounced negative magnetoresistance compared to corresponding two-dimensional (Ga,Mn)As layers. Adopting the concept of the effective magnetic field to magnetotransport allowed us to attribute the angular dependence of the MR largely to the dependence of the NMR on the effective magnetic field. We ascribe the particularly strong anisotropy to strain related effects in the core-shell nanowire, differing topologically from planar structures. For future applications the strong uniaxial anisotropy might be of advantage as it provides two stable magnetization orientations along the nanowire axis.

**Acknowledgements**

We thank T. Wojtowicz, G. Bayreuther, and C. Back for discussions and Matthias Kiessling for performing the SQUID measurements. Support from the Deutsche Forschungsgemeinschaft (DFG) via SFB 689 is gratefully acknowledged. We thank A. Petroutchik and L. T.Baczewski for the preparation of the Au-covered substrates for NW growth.

**Supporting Information:**

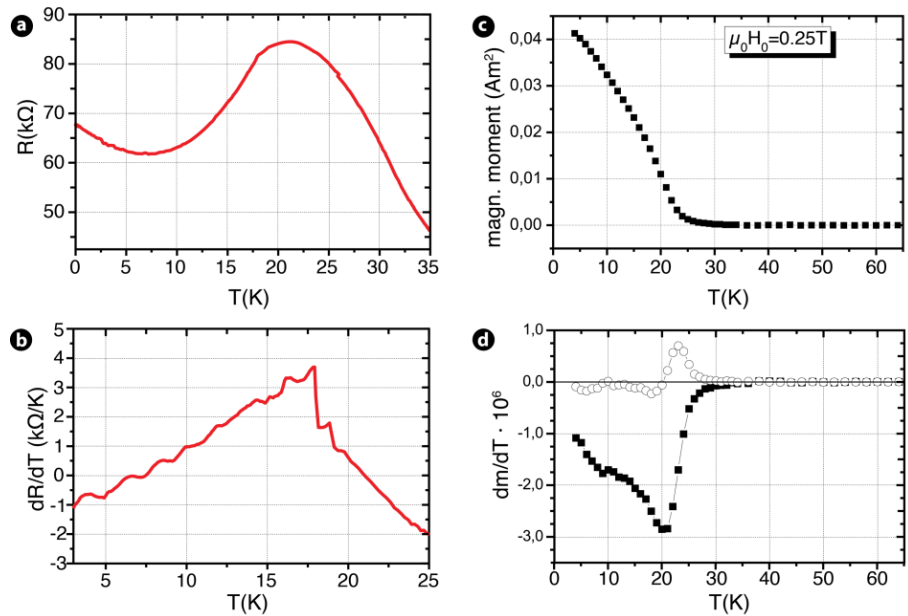

**Figure S1:** Measurement of the Curie-Temperature via magnetotransport (left side) and SQUID (right side). Panel a) displays the resistance of a nanowire as a function of temperature. The decrease of resistance below 20K arises from the appearance of magnetic ordering. The actual Curie-Temperature can be found (following work of Novak et al. [19]) at the singularity in the first derivative of R(T) shown in b). Panel c) displays the magnetic moment $m$ of a nanowire ensemble as a function of temperature. For the measurement an external magnetic field of $\mu_0 H_0 = 250$mT was applied. In d), the first derivative (filled squares) and the second derivative (hollow circles) of m(T) is shown. The measurements reveal a Curie-Temperature of ~18 K (MR-measurement) and ~20 K (SQUID) which are in good agreement.